“On the existence of statistics intermediate between those of Fermi-Dirac and Bose-Einstein“

J. Dunning-Davies,
Department of Physics,
University of Hull,
Hull HU6 7RX,
England.

J.Dunning-Davies@hull.ac.uk

**Abstract.**

Once again the possibility of the existence of particle statistics intermediate between those of Fermi-Dirac and Bose-Einstein surfaces. Here attention is drawn to the fact that some fifteen years ago it was shown that such so-called ‘intermediate’ statistics correspond to no physical process and the stationary probability distributions of intermediate statistics are not compatible with any mechanism which allows a variation between Fermi-Dirac and Bose-Einstein statistics.

“What is between Fermi-Dirac and Bose-Einstein statistics?” This is, at first sight, a seemingly interesting question and maybe that is why it raises its head every few years or so. Whenever this happens, it becomes apparent that people haven’t searched out all the background literature. In 1995, Viefers, Ravndal and Haugset [1] asked this same question and even referred to the ‘fundamental discovery of Leinaas and Myrheim in 1977 of the possibility for intermediate quantum statistics for identical particles in two dimensions interpolating between standard Fermi-Dirac and Bose-Einstein statistics’ [2]. This ignored the fact that such intermediate statistics, with no restriction to two dimensions, had been proposed by Gentile [3] in 1940. As is mentioned in an article of 1989 [4], between 1940 and 1963, several articles appeared which examined both the possibility of systems obeying such intermediate statistics and the actual properties of such systems. However, as was pointed out in1989 [4], the arguments in these various articles seemed to go against the grain of certain limit theorems in probability theory; theorems which assert that, as the number of indistinguishable particles increases without limit while the number of energy cells remains finite, the probability distribution should converge to a normal distribution [5]. However, many authors appeared to feel that it is in precisely this case that the differences between intermediate and other forms of statistic would be manifested. The limit theorems of probability imply that, in the limit, the statistics should become classical!

It has been shown [4] that Bose-Einstein statistics arises *only* in the limit as $d = \infty$, where $d$ represents the number of energy levels. All values of $d$ between one and infinity were shown to give rise to intermediate statistics which cannot be described by stationary probability distributions that tend to the binomial and negative binomial distributions as $d \to 1$ and $d \to \infty$ respectively [6]. Hence, the stationary probability distributions of so-called intermediate statistics are not compatible with any mechanism which allows a variation between Fermi-Dirac and Bose-Einstein statistics. Also, it was realised that no purely thermodynamic argument may be given either in favour of, or against, the existence of intermediate statistics. Therefore, a statistical argument was employed by showing that the stationary probability distribution may only satisfy the recursion relation obtained from the time independent master equation when it coincides with the binomial, negative binomial or Poisson distributions - corresponding to Fermi-Dirac, Bose-Einstein or classical statistics respectively. It seems surprising to realise that this was the first time mention had been made of the Poisson case in studies on intermediate statistics. Classical statistics was shown to emerge as the number of energy cells increases without limit independently of the size of the occupation numbers; it may never depend on the size of the particle number since any conclusions drawn from the generating function and its binomial expansion, in which the particle number merely appears as an index, must lead to the same conclusions. The entire argument appeared to lay to rest the possibility of the existence of particles with spins other than those with semi-integral or integral values.

Note also that it was the basic tenet of the book “Statistical Physics: a Probabilistic Approach” [7] that probability may be used in the development of a statistical basis for thermodynamics. If this is so, it means that any form of physical statistics must be governed by probability distributions, just as fermions and bosons are governed by the negative binomial and binomial distributions respectively. It should be noted that, at high temperatures, these two distributions merge into the Poisson distribution which governs classical, or Boltzmann, particles. However, there is no distribution between the negative binomial and binomial. Hence, it must be concluded that so-called 'intermediate' statistics correspond to no physical process. It should be realised also that the argument is independent of the dimensionality of the system and of any conditions that might have followed from quantum mechanics as Wilczek [8] would have one believe. As far as dimensionality is concerned, it only affects the expression for the density of states, having no effect upon the statistics. The probability distributions govern the statistics of collections of non-interacting particles and so attention is restricted to the negative binomial, binomial and Poisson distributions which govern fermions, bosons and classical particles respectively.

Papers seem to be appearing with monotonous regularity extolling the virtues of these 'intermediate' statistics. Some [9] seem to refer back to the article by Haldane [10] as a basic source. The results obtained mirror those referred to earlier and obtained many years ago in a wide variety of articles; they assume that the occupation number of the energy levels may take any value between one and infinity, perform well-known mathematical manipulations and then make deductions pertaining to physics. These articles are essentially mathematical in nature; physics and statistical theory are not to the fore. One exception is provided by Landsberg [11] whose final conclusion is that there are only two genuine *physical* cases - those described by Fermi-Dirac and Bose-Einstein statistics. This conclusion agrees with the more detailed examination of this problem by Lavenda and Dunning-Davies [4], an article which existed, incidentally, before even that of Haldane.

There may well be physical situations involving apparently non-interacting particles which are not readily explained by Fermi-Dirac or Bose-Einstein statistics but the correct explanations for these particular situations cannot lie with intermediate statistics, at least not until the objections already raised against such statistics [4] are first refuted. Indeed, for the situations falling into this category, it might be worth investigating the possibility that some interaction between the particles is playing a more important role than currently thought. After all, there is no actual physical situation where the particles involved are totally non-interacting. The ideal gases merely provide a convenient, tractable model which may, or may not, accurately reflect physical reality. In some situations, it is the physical model which should be examined critically. Maybe those which continue to spawn discussions of so-called 'intermediate statistics' provide examples where this is so.

**References.**

[1] S.Viefers, F. Ravndal & T. Haugset; *Am. J. Phys.* **63** (1995) 369

[2] J. M. Leinaas & J. Myrheim; *Nuovo Cimento* B **37** (1977) 1

[3] G. Gentile; *Nuovo Cimento* **19** (1940) 493

[4] B. H. Lavenda & J. Dunning-Davies; *J. Math. Phys.* **30** (1989) 1117

[5] R. Von Mises; *Mathematical Theory of Probability and Statistics* (Academic Press, New York, 1964)

[6] B. H. Lavenda; *Int. J. Theor. Phys.* **27** (1988) 1371

[7] B. H. Lavenda; *Statistical Physics: A Probabilistic Approach*, (Wiley, New York, 1991)

[8] F. Wilczek; *Scientific American*, **264** (1991) 24

[9] C. Wolf; *Hadronic J.* **20** (1997) 657
K. Byczuk, et al. arXiv:cond-mat/0403735

[10] F. D. M. Haldane, *Phys. Rev. Lett.* **67** (1991) 937

[11] P. T. Landsberg; *Mol. Phys.* **6** (1963) 341